\documentclass[twocolumn]{biophys}
\usepackage{amssymb,amsfonts,amsthm,bm}
\usepackage[round,numbers,sort&compress]{natbib}
\usepackage[ansinew]{inputenc}

\jno{kxl014}
\gridframe{N}
\cropmark{Y}
\doi{doi}

\begin{document}

\title{Signatures of mechanosensitive gating}

\author{R.~G.~Morris}
\address{National Centre for Biological Sciences, Tata Institute for 
Fundamental Research, GKVK, Bellary Road, Bangalore, 560064, INDIA.}

\begin{abstract}
{
	The question of how mechanically-gated membrane channels open and close is 
	notoriously difficult to address, especially if the protein structure is 
	not available.  This perspective highlights the relevance of 
	micropipette-aspirated single-particle tracking--- used to obtain a 
	channel's diffusion coefficient, $D$, as a function of applied membrane 
	tension, $\sigma$--- as an indirect assay for determining functional 
	behaviour in mechanosensitive channels.  Whilst ensuring that the protein 
	remains integral to the membrane, such methods can be used to identify not 
	only the gating mechanism of a protein, but also associated physical 
	moduli, such as torsional- and dilational-rigidity, which correspond to the 
	protein's effective shape change.  As an example, three distinct $D$ versus 
	$\sigma$ ``signatures'' are calculated, corresponding to gating by 
	dilation, gating by tilt, and a combination of both dilation and tilt. Both 
	advantages and disadvantages of the approach are discussed.
}
{Submitted [DATE], and accepted for publication [DATE].}
{
	*Correspondence: richardgm@ncbs.res.in.\\
	Address reprint requests to Richard G. Morris, National Centre for 
	Biological Sciences, Tata Institute for Fundamental Research, GKVK, Bellary 
	Road, Bangalore, 560064, India.  Tel.:+91 80 2366 6001, ext. 6060.\\
	Editor: [INSERT].
}
\end{abstract}

\maketitle

\markboth{Morris}{Signatures of mechanosensitive gating}

\section*{INTRODUCTION}
The importance of mechanically gated membrane channels was underlined recently 
following the discovery of the {\it Piezo} family of proteins: mammalian 
counterparts to the well studied membrane channels found in invertebrates 
\cite{Coste2010, 
BC+2012,SUK+2012,S-H_W+2014,SSR+2014,SSR+Nat2014,SMC+2015,S-HW+2015}.  However, 
the mechanism by which these channels open and close is not yet fully 
understood.  The state of the art, which has been successful in the study of 
ion channels and pumps, is to infer functional properties from the protein's 
structure, found via X-ray crystallography 
\cite{DAD+98,YJ+02,RD+02,YJ+03,EG_RM_05,SBL+05} or (single particle) 
cryo-electron microscopy \cite{AD_16,JG_+_15}.  Due the complexity of such 
techniques, each has certain limitations.  For the former, crystallizing 
membrane proteins is fraught with difficulty, despite recent advances 
\cite{SBL+07,DA+15,C-YH+15}.  For the latter, resolution can be an issue, 
leading to ambiguity of certain structural details.  In both cases, proteins 
are typically isolated from their natural membrane environment via chemical 
modification \cite{IM+14} (although so-called surfactant-free techniques--- 
whereby a protein is isolated in a co-polymer solvated membrane nanodisc 
\cite{NatProc,dorr}--- are a maturing field which may provide an alternative 
method of purification).  Moreover, once a protein structure has been obtained, 
attempting to infer a gating mechanism is somewhat subjective, particularly in 
the case of mechanosensitive channels, which rely less on the careful 
positioning of localized charge residues than, for example, voltage-gated 
channels.

In the absence of structural data, or to mitigate the aforementioned issues 
with a complimentary assay, an {\it indirect} method is required, where the 
protein exists in its natural environment of a tension-bearing lipid bilayer 
membrane.  For this to be possible, the conformational changes which occur 
during gating must leave a distinct signature in the measurement of another, 
more accessible quantity.  Precisely such a proxy, and part of the associated 
analysis, has already been developed in a different context: quantifying the 
rigidity of a fixed (inactive) conformation protein, KvAP--- a voltage-gated 
potassium channel from {\it Aeropyrum Pernix} \cite{FQ+13,RGM_MST_15}.  By 
providing both the remaining analysis and biological context, this perspective 
aims to assess the broader ramifications of such an indirect assay and its 
potential application for determining functional aspects of mechanically-gated 
channels.  In doing so, concrete calculations are provided that not only 
demonstrate the required theory, but also provide results for experimental 
comparison.

\begin{figure}
\centering
\includegraphics[width=.45\textwidth]{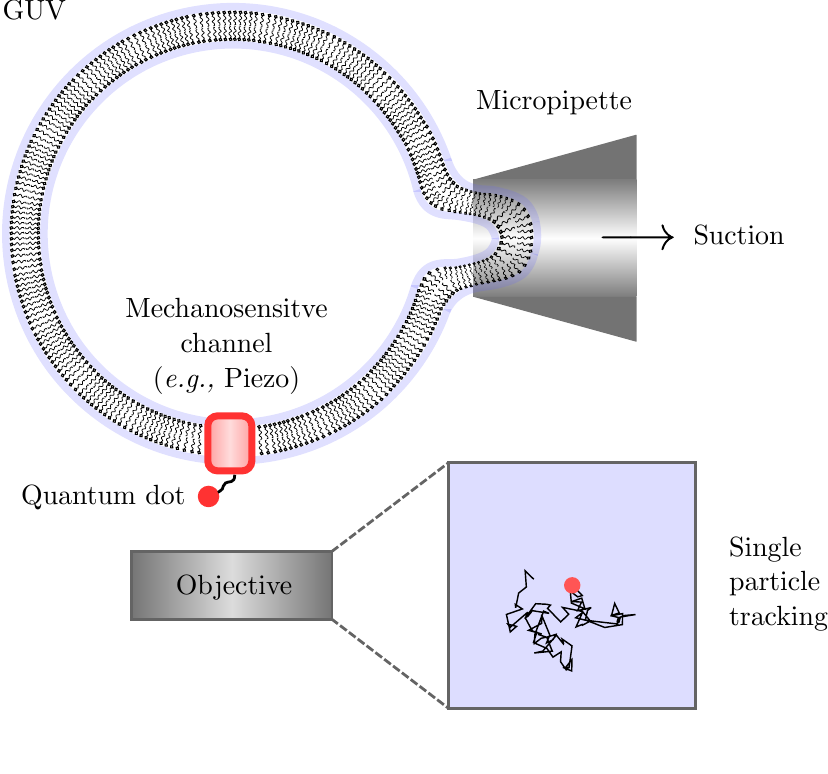}
\caption
{
	Schematic of single particle tracking on the surface of a micropipette 
	aspirated GUV.  As shown in \cite{FQ+13}, this setup can be used to obtain 
	the diffusion constant, $D$, of a membrane channel, as a function of 
	applied tension, $\sigma$.  The resulting $D$ vs.~$\sigma$ plots can be used to 
	identify both the gating mechanism and elastic properties of 
	mechanosensitive channels, such as the {\it Piezo} family of proteins.
}
\label{fig:exp}
\end{figure}

The central idea is that measurements of a protein's diffusion constant, $D$, 
as a function of applied membrane tension, $\sigma$, allow--- via analysis--- 
both conformational changes and effective physical properties, such as elastic 
moduli, to be determined.  The relevant experimental setup is that of 
micropipette-aspirated single-particle tracking, as described in \cite{FQ+13}.  
This involves tracking (via ``quantum dot'' labelling) protein trajectories on 
the surface of a micropipette-aspirated Giant Unilamellar Vesicle (GUV) (see 
Fig.~\ref{fig:exp}) where it is worth remarking that, in principle, the 
purification / reconstitution step might be made surfactant-free by employing 
co-polymner solvated nanodiscs \cite{NatProc,dorr}.  In the context of 
mechanosensitive channels, analysing such data requires that energetic models 
of gating \cite{PW_RP_04,PW_RP_05,RP_TU_PW_PS_09,DR_TU_PS_JK_RP} are combined 
with the classical hydrodynamics of \cite{RGM_MST_15}, where the membrane is 
treated as two-dimensional low Reynolds number fluid.  Here, a protein's 
diffusion coefficient is calculated by first integrating the hydrodynamic 
stresses around the protein boundary to give the drag coefficient, $\lambda$, 
and then linking $\lambda$ to $D$ via the Stokes-Einstein relation 
\cite{batchelor}.

The calculations presented here demonstrate that there are three very distinct 
$D$ versus $\sigma$ relations for three important classes of gating 
\cite{DR_TU_PS_JK_RP}: {\it i}) dilation, where a cylindrically shaped channel 
opens by increasing its radius, {\it ii}) tilt, where opening is achieved by 
deforming from a truncated cone to a cylinder, and {\it iii}) a combination of 
both dilation and tilt.  In the case of {\it i}), the membrane remains planar, 
and the formula of Saffman and Delbr\"{u}ck (SD) may be used, which also takes 
into account the extra drag induced by the three-dimensional embedding fluid.  
By contrast, applying the same protocol in the case of {\it ii}) leads to a 
calculation greatly complicated by geometry.  Here, the conformation of the 
protein and the applied tension determine the shape of the membrane, which in 
turn affects the mobility of the protein via the mechanism of curvature-induced 
shear \cite{MAAD09,MLHAJL10,RGM_MST_15}.  This results in corrections to the SD 
expression, and an entirely different signature of gating.  For case {\it 
iii}), which combines aspects of both pure dilation and pure tilt, the result 
is a surprisingly non-monotonic $D$ versus $\sigma$ curve.

The article concludes by discussing the advantages and disadvantages of 
micropipette-aspirated single-particle tracking as an indirect assay for the 
functional determination of mechanosensitive membrane proteins.

\section*{DILATION}
\begin{figure*}
\centering
\includegraphics[width=1.0\textwidth]{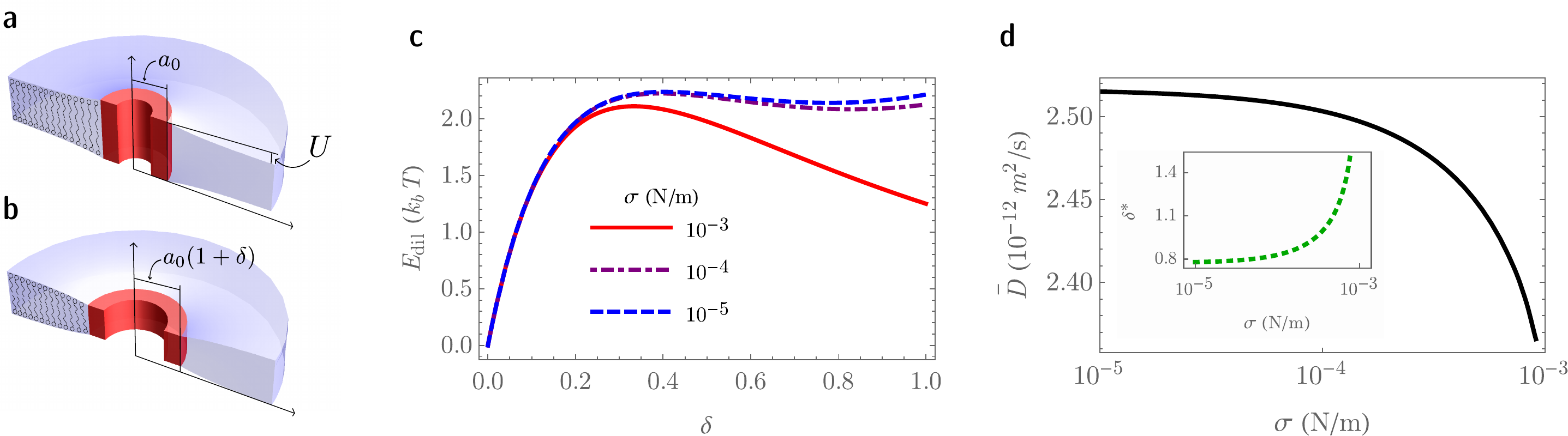}
\caption
{
	Gating by dilation. Panels (a) and (b) are cartoons of a dilation-gated 
	protein channel in closed and open configurations, respectively.  The 
	diagrams indicate the extension (or compression) of the lipid acyl chains 
	in the vicinity of the protein, giving rise to the gating energetics shown 
	in panel (c).  Irrespective of tension, the closed configuration 
	corresponds to $\delta=0$, whilst the open configuration (\textit{i.e.}, 
	the local nonzero minima $\delta^\ast$) is both increasingly stable and 
	occurs at greater dilations [inset of panel (d)], as applied tension is 
	increased.  Panel (d) plots the time average of the diffusion constant for 
	a gating protein in thermal equilibrium (at room temperature) as a function 
	of applied tension.  The decrease of $\bar{D}$ at higher tensions is due to 
	open configurations having larger radii and being more stable.
}
\label{fig:dil}
\end{figure*}

As a cylindrically-shaped channel opens via dilation, its diffusion constant, 
$D$, decreases, due to the increase in the protein's radius, $a$.  This can be 
seen from the SD formula,
\begin{equation}
	D^{\left( \mathrm{SD} \right)} = \frac{k_B T}{4\,\pi\,\eta}\left[ 
	\log{\left(\frac{\eta}{\mu\,a}\right)} -\gamma\right], \label{eq:SD}
\end{equation}
which treats the membrane as a two-dimensional fluid of viscosity $\eta$, 
embedded in a three-dimensional fluid of viscosity $\mu < \eta$ (both at low 
Reynolds number).  Here, $k_B$ is Boltzmann's constant, $\gamma$ is Euler's 
constant, $T$ is temperature, and the effects of vesicle curvature are 
neglected since the radius of a GUV ($\mu\mathrm{m}$) greatly exceeds that of a 
membrane-protein ($\mathrm{nm}$). 

The relationship between protein radius, $a$, and membrane tension, $\sigma$, 
is given by the energetics of gating, assumed here to be that of ``hydrophobic 
mismatch'' \cite{PW_RP_04,PW_RP_05,RP_TU_PW_PS_09}.  The approach characterises 
the free energy of the (static) membrane plus protein system with just two 
terms.  The first term $\pi\,\mathcal{K}\,(a-a_0)\,d^2$ arises from assigning a 
Hookean elastic energy to the lipids that border the protein, which either 
extend or compress their acyl chains (by an amount $d$) in order to match the 
``height'' of the hydrophobic region on the exterior of the protein.  Since 
experiments indicate the hydrophobic height decreases as $a$ increases, a 
simple volume conserving model is adopted, implying $d=a_0^2\,U/a^2$, where 
$a_0$ is just the lower bound on protein radius {\it i.e.}, $a\in [a_0, 
	\infty)$.  (The constant $U$ is then just the extension of the lipid acyl 
	chains when $a=a_0$).  The second term arises from the competition between 
	the elastic deformation of the channel and the surface tension $\pi 
	(k-\sigma)(a-a_0)^2$.  Up to an additive constant, the combined energy is 
	given by\footnote{Notice that, rather than invoke a term corresponding to 
		steric hindrance at small radii, it suffices to only consider energies 
	greater than some value $E(0)$, since $\delta=0$ is always a global minimum 
of the function, by construction.}
\begin{equation}
	E_{\mathrm{dil}} = 
	\frac{\pi\,\mathcal{K}\,a_0\,\delta\,U^2}{(1+\delta)^4} + \pi 
	(k-\sigma)\,a_0^2\,\delta^2,
\label{eq:E_dil}
\end{equation}
which has been written using a dimensionless dilation $\delta = (a/a_0) - 
1$.  Estimates for $\mathcal{K}$, $U$ and $a_0$ can all be taken from the 
literature (see Numerical Values section), whilst the elastic constant $k$ is 
chosen to be the minimum value which ensures that (\ref{eq:E_dil}) is bounded 
for all physically plausible tensions $\sigma < 10^{-3}$ N/m (the membrane of a 
GUV is know to tear at tensions of around $10^{-3}$ N/m).  Imposing the 
constraint $\partial E_{\mathrm{dil}} / \partial\delta = 0$ gives rise to two 
minima, located at $\delta=0$ and $\delta=\delta^\ast(\sigma)$, which are taken 
as ``closed'' and ``open'' states, respectively.  A schematic of these states 
is depicted in Figs.~\ref{fig:dil}(a) and \ref{fig:dil}(b), whilst 
Fig.~\ref{fig:dil}(c) and the inset of Fig.~\ref{fig:dil}(d) demonstrate that 
the position and depth of the local minimum increases with applied tension.

Assuming that the channel is in equilibrium with the environment, the average 
rate at which thermal fluctuations cause the channel to open is just the escape 
rate, $R_{\delta=0}$, from the energy minimum at $\delta=0$, and is given by 
Kramer's formula \cite{Gardiner,NGVK}: $R = \mathcal{C}\,\exp\left[ 
-E_{\mathrm{barrier}}/k_B\,T \right]$, where $\mathcal{C}$ is a constant and 
$E_{\mathrm{barrier}} = \max_{\delta\in[0,\delta^\ast]} E(\delta)$.  Similarly, 
the rate at which fluctuations cause the channel to close, 
$R_{\delta=\delta^\ast}$, is given by $R_{\delta=\delta^\ast} = 
\mathcal{C}\,\exp\left[ 
-\left(E_{\mathrm{barrier}}-E(\delta^\ast)\right)/k_B\,T \right]$.  The ratio 
of these rates is just the ratio of mean first passage times for traversing the 
energy barrier.  Therefore, over long times, the fraction of time spent at 
$\delta=0$ and $\delta=\delta^\ast$ approaches $1/\left[ 1+\exp\left( -\Delta 
E/k_B\,T \right)\right]$ and $1/\left[ 1+\exp\left( \Delta E/k_B\,T 
\right)\right]$, respectively, where $\Delta E = E(\delta^\ast) - E(0)$.  The 
effective diffusion constant, $\bar{D}$, is calculated by substituting 
$\delta=0$ and $\delta=\delta^\ast$ into (\ref{eq:SD}) and then taking an 
average, weighted by the aforementioned time fractions.  The result 
[Fig.~\ref{fig:dil}(d)] indicates that the diffusion constant of a 
dilation-gated channel decreases with applied tension, but only on a small 
scale ($\sim 10^{-13}$ m$^2$/s over three orders of magnitude of $\sigma$).

\section*{TILT}
\begin{figure*}
\centering
\includegraphics[width=1.0\textwidth]{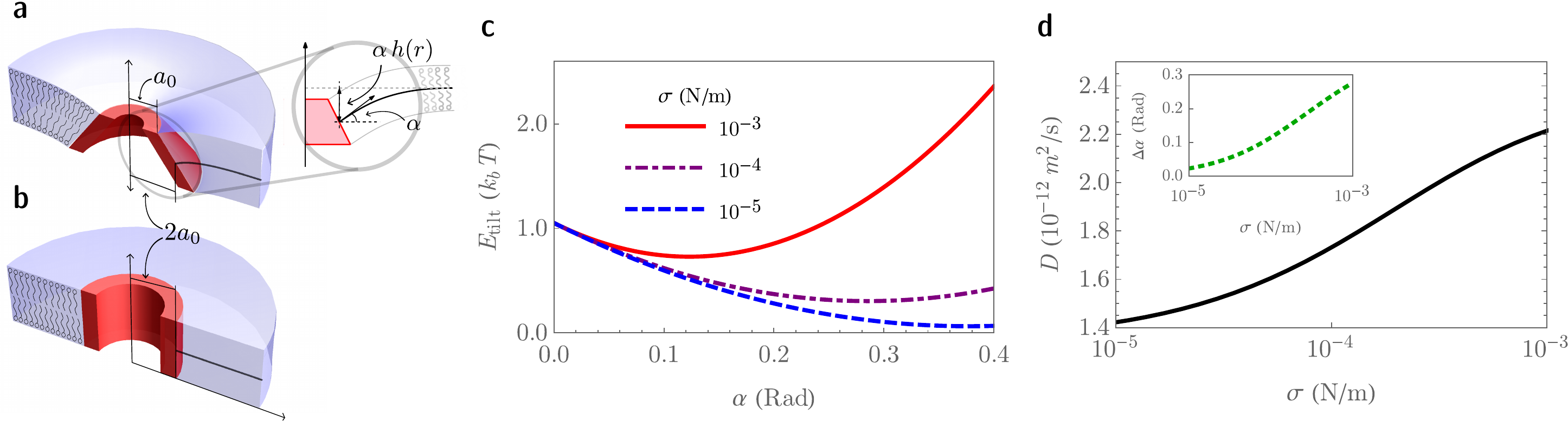}
\caption
{
	Gating by tilt.  Panels (a) and (b) are cartoons of a tilt-gated protein 
	channel at low and high tensions, respectively.  The energy of the protein 
	as a function of tilt angle $\alpha$ [panel (c)] has a single minima, 
	$\alpha^\ast$, where the torque due to applied tension is balanced by the 
	protein's torsional rigidity.  As tension, $\sigma$, is increased, 
	$\alpha^\ast$ decreases and the channel becomes increasingly cylindrical.  
	Panel (d) shows that the diffusion constant of a tilt-gated protein 
	increases with tension.  This is because the large tilt-angles which 
	correspond to low tensions imply a large curvature-induced shear, and hence 
	low mobility.  At higher tensions, the diffusion constant saturates as the 
	protein becomes effectively cylindrical.  The inset of panel (d) 
demonstrates that the protein undergoes substantial angular strains across the 
range of experimentally accessible tensions.}
\label{fig:tilt}
\end{figure*}
In contrast to gating via dilation, gating via tilt involves channels in the 
shape of a truncated cone, which deform the membrane in the vicinity of the 
protein [see Figs.~\ref{fig:tilt}(a) and \ref{fig:tilt}(b)] and lead to 
corrections to (\ref{eq:SD}).  This is an example of so-called 
curvature-induced shear: the phenomenon whereby Gaussian curvature of the 
membrane modifies shear and hence drag and diffusion 
\cite{MAAD09,MLHAJL10,RGM_MST_15}.

In order to calculate such corrections, it is therefore necessary to calculate 
the membrane shape induced by the channel.  In principle, this shape not only 
depends on the conformation of the channel, but also its motion.  That is, a 
non-zero channel velocity results in a spatially varying two-dimensional 
pressure ({\it i.e.}, tension), as well as a rotation-inducing torque on the 
channel, breaking the angular symmetry of deformations.  However, in practice, 
since the drag itself is the leading order coefficient in a power series 
expansion of force in terms of velocity, such effects can be neglected when 
calculating $\lambda$, and the symmetric, equilibrium, shape suffices.  Here, 
the equilibrium mid-surface, denoted $\mathcal{S}$, is calculated by minimising 
a Helfrich-like free energy functional \cite{WH73}, 
\begin{equation}
	E_{\mathrm{mem}}=\int_\mathcal{S} \left(2\kappa\,H^2 + \bar{\kappa}\,K + 
	\sigma\right)dA,
	\label{eq:E}
\end{equation}
which describes a bilayer under surface tension $\sigma$ that also has a 
bending energy (per unit area) of $2\kappa\, H^2 + \bar{\kappa}\,K$, where 
$\kappa$ and $\bar{\kappa}$ are constant bending moduli and $H$ and $K$ are the 
mean and Gaussian curvatures, respectively. Using a small angle approximation, 
the solution to the corresponding Euler-Lagrange equation can be characterised 
by an axisymmetric height field $\alpha\,h(r),\ \forall\  r\in [a,\infty)$, 
	where $\alpha$ is the tilt-angle, subtended at the ``walls'' where the 
	channel meets the membrane, and the function $h$ is given by 
	\cite{RGM_MST_15,TRW_MMK_WH_98}
\begin{equation}
	h(r) = \frac{\ell\,K_0\left( r/\ell \right)}{K_1\left(  a/\ell\right)}.
	\label{eq:h}
\end{equation}
Here, $K_n$ represents an order-$n$ modified Bessel function of the second kind 
and the characteristic length scale is given by $\ell=\sqrt{\kappa/\sigma}$.  
The boundary terms of the same calculation yield the torque applied at the 
channel walls, between the membrane and protein \cite{RGM_MST_16}
\begin{equation}
	\tau = 2\pi\,\alpha\,\left(a\,\sigma\,h(a) - \bar{\kappa}\right),
\label{eq:tau}
\end{equation}
which is assumed to be hinged about the mid-plane of the membrane [see 
Figs.~\ref{fig:tilt}(a) and \ref{fig:tilt}(b)].  If a conformation is to be 
stable, (\ref{eq:tau}) must be balanced by the protein's torsional rigidity, 
implying that the gating energy takes the form
\begin{equation}
	E_{\mathrm{tilt}} = \pi\,\left[\sigma\,a\,h(a) + 
	\bar{\kappa}\right]\alpha^2 - \frac{k^\prime}{2}\alpha^2 + 
	\left[\tau_{\mathrm{ref}}\left(\sigma_{\mathrm{ref}}\right) - 
	k^\prime\,\alpha_{\mathrm{ref}}\right]\,\alpha,
	\label{eq:E_tilt}
\end{equation}
where the first term can be deduced from (\ref{eq:tau}), the second term has 
the form of a torsion spring with rigidity $k^\prime$, and the third term is 
just the known net torque at some reference angle, $\alpha_{\mathrm{ref}}$, and 
reference tension, $\sigma_{\mathrm{ref}}$.  By analogy with the case of 
dilation, the relationship between $\alpha$ and $\sigma$ is provided by 
imposing $\partial E_{\mathrm{tilt}}/\partial\alpha = 0$.  This leads to a 
single minimum, whose depth, and position, $\alpha=\alpha^\ast(\sigma)$, 
increase as applied tension decreases [Fig.~\ref{fig:tilt}(c)].

What remains is to calculate the diffusion constant, where the standard 
approach of classical hydrodynamics may be used, {\it i.e.}, by solving the 
equations of incompressible Stokes flow in order to calculate the drag and then 
invoking the Stokes-Einstein relation.  However, due to the geometry of the 
membrane, the hydrodynamics must now be formulated in a covariant way.  That 
is, local quantities are expressed in the tangent plane of the membrane 
mid-surface $\mathcal{S}$ [{\it i.e.}, defined by $\alpha^\ast\, h(r)$].  In 
general, such covariant Stokes equations are very hard to solve in all but a 
handful of special cases.  However, in this case, the small size of angle 
$\alpha^\ast$ may be exploited, and therefore approximate solutions can be 
obtained as power series in $\alpha^\ast$.  This type of perturbative scheme is 
explained in detail in \cite{RGM_MST_15} (and the associated supplementary 
material), where the covariant Stokes flow problem is solved by moving to a 
description in terms of a scalar stream function.  After separating-out the 
angular dependence, the result is a scheme of fourth-order ODEs and boundary 
conditions, each of which corresponds to the radial part of a single 
coefficient in the perturbative expansion of the stream function in terms of 
$\alpha^\ast$.  The zeroth order contribution is constructed so that the SD 
result is recovered as $\alpha^\ast\to 0$ where, since the characteristic 
membrane length scale $\ell$ is much less than the SD length $\eta/\mu$ (true 
for tensions $\sigma \ge 2\times 10^{-7}$ N m$^{-1}$) the role of the embedding 
fluid may then be ignored in all higher order corrections.  The next lowest 
order correction is at $(\alpha^\ast)^2$, since diffusion cannot depend of the 
sign of $\alpha^\ast$ due to the up/down symmetry of the membrane.  Obtaining 
the coefficient of this correction involves solving the radial part of an 
inhomogeneous biharmonic equation, the particular solution of which must be 
computed numerically.  Despite the lack of a closed form result, the numerics 
can still be used to calculate corrections to Cauchy stress tensor and 
therefore both the drag and diffusion constants.  Without recapitulating the 
detailed calculation of \cite{RGM_MST_15}, the result is that  
\begin{equation}
	D=D^{(\mathrm{SD})} + (\alpha^\ast)^2\,D^{(2)} + O\left[ (\alpha^\ast)^4 
	\right],
	\label{eq:D_expand}
\end{equation}
\begin{figure*}[!t]
\centering
\includegraphics[width=1.0\textwidth]{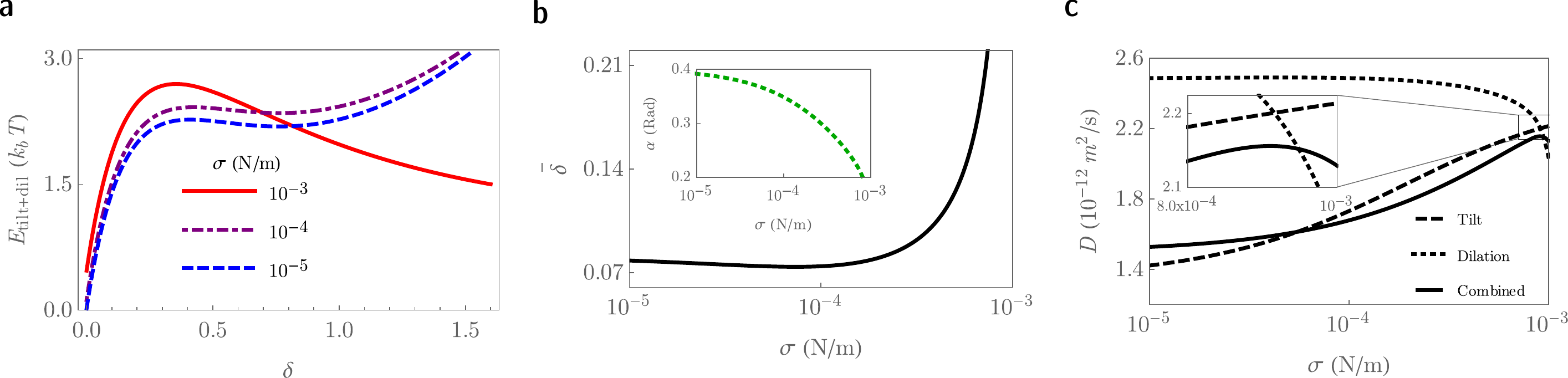}
\caption
{
	Gating by combined tilt and dilation.  The energy has two minima, 
	indicating distinct open and closed states [panel (a)].  The behaviour of 
	the channel (both dilation and tilt) under applied tension is shown in 
	panel (b), demonstrating the three regimes identified in the main text: 
	{\it i}) tilt combined with contraction, {\it ii}) tilt dominated, and {\it 
	iii}) dilation dominated.  Panel (c) demonstrates the three unique 
	signatures ($D$ vs. $\sigma$ plots) corresponding to pure dilation, pure 
	tilt, and combined dilation and tilt.  The magnified inset clarifies that 
	the combined signature is non-monatonic at high tensions.
}
\label{fig:comb}
\end{figure*}
where the numerically calculated coefficient $D^{(2)}$ is independent of 
$\alpha^\ast$ but still relies on the both $\sigma$, $h(r)$, and its higher 
derivatives.  The result is plotted in Fig.~\ref{fig:tilt}(d), and indicates 
that as tension increases, the angular strain and hence diffusion constant 
increases until saturation when the protein effectively becomes cylindrical.  
The heuristic is that low tensions imply large tilt-angles, and therefore 
membrane shapes with large Gaussian curvature at the boundary of the protein.  
Such large curvatures induce large shears in the fluid flow, increasing drag 
and reducing protein diffusion.   

\section*{COMBINED DILATION AND TILT}
The energetics of combined dilation and tilt is assumed to be the sum 
$E_{\mathrm{dil}+\mathrm{tilt}} = E_{\mathrm{dil}}+E_{\mathrm{tilt}}$, where 
the substitution $a=a_0(1+\delta)$ has been made in $E_{\mathrm{tilt}}$.  The 
result is a function of both $\alpha$ and $\delta$, which is minimised by the 
constraints $\partial E_{\mathrm{dil}+\mathrm{tilt}} / \partial \alpha = 0$ and 
$\partial E_{\mathrm{dil}+\mathrm{tilt}} / \partial \alpha = 0$.  Solving first 
for $\alpha^\ast(\delta,\sigma)$, substituting into 
$E_{\mathrm{dil}+\mathrm{tilt}}$, and plotting the result against $\delta$ 
gives rise to Fig.~\ref{fig:comb}(a).  As in the case of pure dilation, there 
are two minima, zero and $\delta^\ast(\sigma)$ which represent ``closed'' and 
``open'' states, respectively.  Taking the average of these values, weighted by 
the fraction of time that a protein in thermal equilibrium would spend in each 
state, gives rise to Figs.~\ref{fig:comb}(b) and \ref{fig:comb}(b) inset.  
Here, the behaviour of the system can be characterised by roughly three 
regimes:  {\it i}) Low tensions ($\sim 10^{-5}$ N/m), where the protein is 
conical and the response of the system to increasing tension only gives rise to 
small decreases in $\alpha$ which, in order to accommodate the changing 
membrane shape, is compensated-for by small contractions in the channel radius.  
{\it ii}) Mid-range tensions ($\sim 10^{-4}$ N/m), where the channel radius is 
effectively constant, and the response to increasing tension is to reduce the 
tilt-angle causing the channel to become increasingly cylindrical.  {\it iii}) 
High tensions ($\sim 10^{-3}$ N/m), where the tilt-angle is very small, and 
pure dilation is recovered, {\it i.e.}, increasing tension increases the radius 
of an essentially cylindrical channel.

The resulting diffusion constant is shown by the solid black line of 
Fig.~\ref{fig:comb}(c), where the lines for pure dilation (dotted) and pure 
tilt (dashed) are also shown.  For the case of ``tilt + dilation'', at low 
tensions, the combined effects of decreasing tilt-angle and radial contraction 
almost cancel each other out leading to a very shallow, nearly flat, $D$ vs.  
$\sigma$ curve.  In the mid-range of tensions, the curve begins to rise 
steeply, and resembles the pure tilt curve, shifted along the horizontal axis.  
Finally, at high tensions, the protein is almost cylindrical and therefore 
increasing the tension leads to pure dilation.  The result is that the curve 
peaks and diffusion begins to decrease as tension increases 
[Fig.~\ref{fig:comb}(c), magnified section].

\section*{DISCUSSION}
The calculations in this article demonstrate that different types of gating can 
be adequately distinguished by their respective $D$ versus $\sigma$ plots.  
Moreover, given data, the approaches set out here could feasibly be used to 
predict effective elastic moduli, such as the torsional- or radial-rigidity, by 
curve fitting, as in \cite{RGM_MST_15}.  However, that study concerned KvAP in 
its inactive conformation, and although mechanical forces have been suggested 
to modulate the voltage at which gating occurs, the method is most relevant to 
{\it mechanosensitive} channels, such as the aforementioned Piezo family, for 
example, whose {\it function} is directly related to tension.  

However, correctly applying such methods to real-world proteins presents 
certain challenges.  For example, single-particle cryo-EM indicates that the 
shape of Piezo1 is akin to a ``propeller'' \cite{JG_+_15}, thus breaking 
angular symmetry and greatly complicating the aforementioned calculations.  
Similarly, such proteins exhibit certain unique traits, such as relatively 
rapid rates of inactivation \cite{Coste2010}, for example.  These features must 
either emerge from a full analysis, or be taken account-of by the gating 
energetics.

In the context of detailed crystallographic approaches, the methods proposed 
here should be seen as a complimentary assay for investigating conformational 
changes in a coarse-grained way.  That is, the approach overlooks complicated 
internal re-arrangements, focussing instead on a channel's {\it effective} 
shape and {\it effective} elastic properties.  Indeed, since only a diffusion 
coefficient needs to be measured, some of the technical issues associated with 
structural data may be sidestepped: neither crystallization nor chemical 
modification of the channel are required, and elastic moduli can be estimated 
directly by fitting gating models to the data.  We note that the experimental 
accuracy of measuring diffusion coefficients via repeated single-particle 
tracking is around $\pm 0.1\times 10^{-12}$ m$^2$/s \cite{FQ+13}, and therefore 
properly resolving different gating mechanisms requires, as a minimum, 
gathering data over tensions ranging three orders of magnitude.  Furthermore, 
the interesting features of combined gating occur at very high tensions ({\it 
circa} $10^{-3}$ m$^2$/s), above which bi-layer membranes are known to tear.  
Indeed, such issues notwithstanding, the approach still relies on a judicious 
choice of model for fitting and therefore, ideally, information from both 
coarse-grained and crystallographic approaches should be {\it combined} in 
order to fully understand membrane channel gating.

Going forwards, although this perspective argues that such techniques offer 
great promise, more work is clearly needed in order to refine certain aspects, 
both in the presence and absence of crystallographic data.  Possible extensions 
may include the effects of membrane composition and annular lipid-protein 
interactions, or membrane curvature, neither of which were considered here.  It 
is also worth remarking that, so far, proteins have only been treated in 
isolation, which is achieved experimentally by using very low concentrations.  
However, in the cell, concentrations of membrane-bound proteins are likely to 
be much higher, leading to cooperative gating effects, such as those described 
in \cite{Haselwandter2013}. One might speculate that such behaviour translates 
into spatial correlations between particle trajectories and hence modifications 
to diffusion and its corresponding $D$ versus $\sigma$ signature.  Finally, it 
may also be possible to measure the functional response of a channel alongside 
its diffusion coefficient.  That is, to monitor pH changes, or trans-membrane 
potentials, in order provide evidence that a channel is working or not.  Either 
way, further work in the area is welcome. 

\section*{NUMERICAL VALUES}
The hydrodynamic viscosity of the membrane ($\eta=6\times 
10^{-10}\,\mathrm{kg}\,\mathrm{s}^{-1}$) and surrounding fluid 
($\mu=10^{-3}\,\mathrm{kg}\,\mathrm{m}^{-1}\,\mathrm{s}^{-1}$) were taken 
from \cite{FQ+13}.  The minimum channel radius ($a_0 = 2\times 
10^{-9}\,\mathrm{m}$), maximum acyl-chain extension 
($U=10^{-10}\,\mathrm{m}$), and corresponding spring constant 
($\mathcal{K}=4\times10^{9}\,\mathrm{J}$), were taken from \cite{PW_RP_05}.  
The elastic moduli of the lipid bilayer ($\kappa = 20\, k_B\, T$, and 
$\bar{\kappa} = 0.9\,\kappa$, for $T=298\,\mathrm{K}$) were taken from 
\cite{MH_JJB_MD_12}.  The torsional rigidity ($k^\prime=100\,k_B\,T$) was 
taken to match the order of magnitude of the voltage-gated channel KvAP 
\cite{RGM_MST_15}.  Dilational rigidity 
($k=10^{-3}\,\mathrm{J}\,\mathrm{m}^{-2}$) was chosen to be the minimum 
value that ensured the energy was bounded for tensions below 
$10^{-3}\,\mathrm{N}\,\mathrm{m}^{-1}$.  Similarly, the reference tilt 
angle ($\alpha_{\mathrm{ref}} = 0.4\,\mathrm{Rad}$) at very low tension 
($\sigma_{\mathrm{ref}}=10^{-8}\,\mathrm{N}\,\mathrm{m}^{-1}$) was chosen 
to correspond to the maximum tilt found in KvAP \cite{RGM_MST_15}.  



\section*{ACKNOWLEDGMENTS}

\ack{
The author acknowledges both comments from and discussions with: 
Prof.~M.~S.~Turner (Warwick, United Kingdom), Prof.~M.~Rao (NCBS, India), 
Dr.~A.~Rautu (NCBS, India), and K.~Husain (NCBS, India).  The author thanks the 
Tata Institute for Fundamental Research (India) and the Simons Foundation (USA) 
for financial support.
}

\end{document}